\documentclass[12pt,english]{article}
\pdfoutput=1
\usepackage[T1]{fontenc}
\usepackage[latin1]{inputenc}
\usepackage{array}
\usepackage{graphicx}
\usepackage{setspace}
\usepackage[authoryear]{natbib}

\makeatletter

\providecommand{\tabularnewline}{\\}

\usepackage{xspace}

\newcommand{\Ang}[1]{${#1}$\AA\xspace}

\date{}

\usepackage{babel}
\makeatother
\begin{document}

\title{Identification of specificity determining residues in enzymes using
environment specific substitution tables}

\author{Swanand Gore and Tom Blundell\\
\{swanand,tom\}@cryst.bioc.cam.ac.uk\\
Department of Biochemistry, University of Cambridge\\
Cambridge CB2 1GA England}

\maketitle
\begin{abstract}
Environment specific substitution tables have been used effectively
for distinguishing structural and functional constraints on proteins
and thereby identify their active sites (\citet{distinguish_str_func_restr}).
This work explores whether a similar approach can be used to identify
specificity determining residues (SDRs) responsible for cofactor dependence,
substrate specificity or subtle catalytic variations. We combine structure-sequence
information and functional annotation from various data sources to
create structural alignments for homologous enzymes and functional
partitions therein. We develop a scoring procedure to predict SDRs
and assess their accuracy using information from bound specific ligands
and published literature.\newpage

\end{abstract}

\section{Introduction}

Enzymes are critical to cellular machinery. Enzymes are believed to
have developed different specificities following gene duplication
events that ease the evolutionary pressure on copies and allow exploration
of novel avenues to greater organismal fitness. Each copy then develops
its own niche, characterized by expression and localization, catalytic
mechanism, substrate specificity, cofactor dependence and catalysis
products. Such paralogous enzymes should have an evolutionary imprint
corresponding to their specific niche, in addition to maintenance
of structural fold. Thus evolutionary analysis of available structural
and sequnce data should enable identification of key residues responsible
for specificity of various kinds. Enzyme specificity can be estimated
with functional assays without structure determination, but identification
of SDRs (specificity determining residues) remains difficult. While
ENZYME (\citet{ENZYME}) - a database of enzyme sequences with detailed
functional annotation - exists, there is no such database of SDRs.
Time, cost and technical limitations slow down structure determination
and even when structure is known, it is not trivial to identify the
residues important for binding cofactors and substrates. Hence it
is important to be able to identify such residues computationally.
Reliable detection of such residues will aid in deciding whether a
SNP is deleterious or neutral and suggest mutation studies. Function
assignment to sequence could be done at a finer level, e.g. by verifying
that SDRs necessary for certain substrate are present. Computational
SDR identification has received a lot of attention and several methods
have been proposed. Evolutionary trace (ET) is one of the most important
methods (\citet{evoltraceStrClust}, \citet{EThybridMethods}). It
builds a phylogenetic tree based on sequence comparisons, such that
branch lengths are indicative of evolutionary divergence. Functional
subgroups consist of sequences in subtrees determined from this tree
using a divergence cutoff. Residues common to a subtree are considered
specificity-conferring rather than the ones common to entire tree.
Spatial cluster identification can be used with ET to reduce the number
of false positives. Inferring phylogeny correctly remains the main
cause of concern in this approach, hence attempts have been made to
use existing annotation with various statistical techniques. Another
important direction is to use spatial proximity of residues.

Cornerstone of our approach is that structural environment influences
residue substitution patterns, illustrated by \citet{earlyESST} and
later used effectively for structure-sequence alignment and fold recognition
(\citet{fugue}). Structural environment of a residue is described
in terms of secondary structure, solvent accessibility, sidechain-sidechain
and sidechain-mainchain hydrogen bonding. Residue substitution tables
derived from a set of high quality sequence-structure alignments represent
the expected substitution rate in a structural environment. Unexpected
conservation of a residue is indicative of functional restraint acting
on it. Advantage of using ESSTs is that the structurally conserved
residues are masked, which is why active sites of homologous enzymes
can be identied reliably with this approach. This approach has been
extended in the present work by using functional annotation information.

A set of homologous enzymes is generally a union of smaller functionally
specific subsets, e.g. substrate-specific subsets in serine proteinases
(trypsin, chymotrypsin etc.), cofactor-specific subsets in ferrodoxin
reductases (NAD and NADP specific) and so on. In multiple sequence
alignment of a homologous protein family, SDRs generally appear as
differentially conserved subcolumns. But all such appearances would
not be SDRs. Our hypothesis is that SDRs would be identified by combining
differential conservation with ESST-based detection of functional
restraint.

\section{Families, functional partitions and profiles}

\begin{figure}

\caption{Workflow}

\begin{center}\includegraphics[%
  width=150mm]{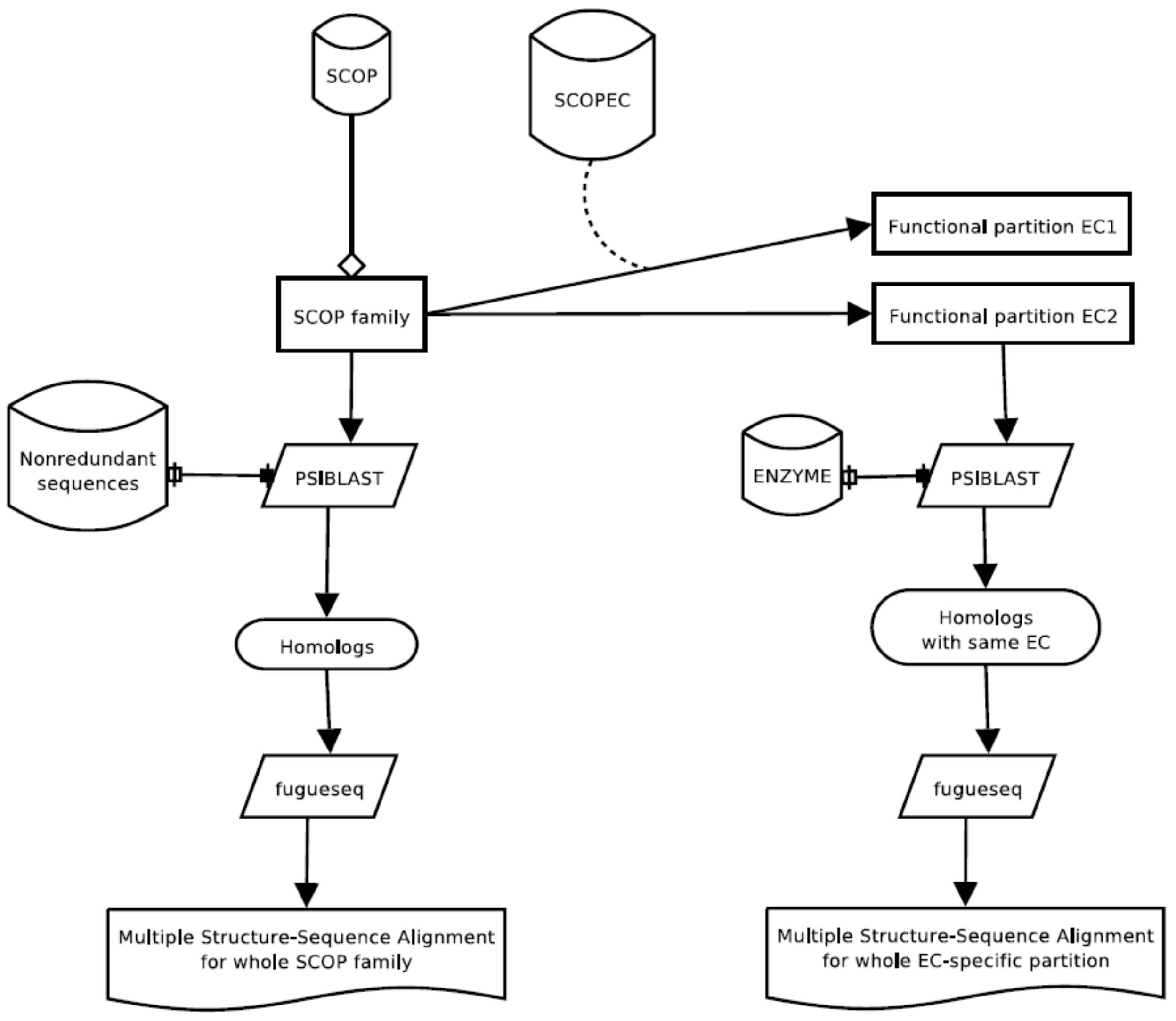}\end{center}

\label{workflow}
\end{figure}

In order to test our hypothesis, we need to construct a dataset of
homologous enzyme families with reliable functional partitions in
them. While SCOP classification can be used in a straightforward way
for making families, identifying functionally specific subsets is
not a trivial task. Some automated approaches to detect functional
shift, e.g. \citet{funshiftakker}, exist to infer such partitions
but manual annotation remains the most reliable. Additionally, protein
function is not a precise and quantifiable entity. This restricted
our study to enzymes which are the the most well studied and well
annotated class of proteins. Enzyme function is fairly well defined
and well classified according to hierarchical Enzyme Classification
scheme (EC). We use the mapping between SCOP domains and EC numbers
(\citet{scopec}) to make EC-specific subgroups within a SCOP domain
family. We generate profiles (multiple structure-sequence alignments)
for SCOP families and functional partitions. Sequence homologs for
structural families were found using PSIBLAST (\citet{psiblast})
on nonredundant sequence database, whereas function-specific partitions
were enriched using PSIBLAST searches on ENZYME database (\citet{ENZYME}).
PSIBLAST hit on ENZYME database is retained only if the EC number
of hit matches that of query. All PSIBLAST searches were with 5 rounds
and e-value 0.01, hits smaller than 75\% of query length were ignored.
All structure-sequence alignments were carried out with fugueseq (\citet{fugue})
which has been shown to improve alignment quality over PSIBLAST. This
process is summarized in Fig.\ref{workflow}.

Another constraint on the choice of dataset comes from the need for
sufficient functional diversity in a SCOP domain family. In its absence,
the contrast between the domain family and EC-specific subgroup within
it might not be detectable. Hence we chose the SCOP families with
at least two different EC annotations.

To be able to test the hypothesis quantitatively, a gold standard
set of SDRs for every enzyme is needed. But SDRs are generally a topic
of lively debate among researchers, partly due to the infeasibility
of performing all necessary mutation studies. Thus there is no such
dataset in our knowledge. Hence we use the information of bound ligands
and close-by residues to assess the hypothesis. Due to this, the dataset
gets restricted to only those cases where at least one EC-specific
domain group has a relevant ligand bound. A relevant ligand is the
one unique to the reaction carried out by that EC-group among all
possible reactions in that domain family. For example, in SCOP family
c.1.10.4 there are two functional subgroups:

3-deoxy-8-phosphooctulonate synthase (EC 2.5.1.55) : Phosphoenolpyruvate
+ D-arabinose 5-phosphate + H(2)O = 2-dehydro-3-deoxy-D-octonate 8-phosphate
+ phosphate

3-deoxy-7-phosphoheptulonate synthase (EC 2.5.1.54) : Phosphoenolpyruvate
+ D-erythrose 4-phosphate + H(2)O = 3-deoxy-D-arabino-hept-2-ulosonate
7-phosphate + phosphate

Here D-arabinose 5-phosphate is unique to EC 2.5.1.55 and is present
in domain 1fxqA as A5P. Hence it is taken as an indicator of SDR locations
and not phosphienolpyruvate which is common cofactor in both reactions.
We sometimes use products also as such indicators. Ligand is considered
relevant if its name from the PDB file (HETNAM, HETSYM records) matches
its name in the reaction or PDBsum (\citet{pdbsum}) finds it sufficiently
similar to ideal ligand molecule. Our final dataset consists of 97
examples drawn from 68 families. Very few SDR identification studies
are carried out with these many examples.

\section{Profiles and substitution patterns}

Structural and sequence information in MSSA can be misleading if dominated
by very close homologs, hence each MSSA was filtered with 90\% sequence
identity cutoff to avoid redundancy.

Observed substitution pattern for a column in profile MSSA (multiple
structure-sequence alignment) was calculated after weighing down contributions
from similar sequences ($>60\%$ sequence identity). Gaps were ignored
while calculating the observed substitution pattern but the ratio
of gaps to amino acids in a column was computed. Columns with high
gap content are generally not functional hence gap content was used
as a filtering criterion as described later. Observed substitution
patterns are normalized and sequence entropy was also calculated to
get a measure of variability in the column as $\sum_{i=1}^{20}-f_{i}log(f_{i})$,
where $f_{i}$ is the fraction of $i^{th}$ amino acid in the distribution.

Expected substitution patterns for a column were calculated using
environment specific substitution probability tables derived from
high quality multiple structure alignments from 371 families (\citet{fugue}).
Substitution probabilties from every structure were averaged to get
expected substitution probabilities for each column in MSSA. Again,
sequence-based clustering was used to avoid expected substitution
pattern getting dominated by very similar structures.

Functional restraint is calculated as the city-block distance between
normalized observed and predicted substitution patterns ($\sum_{i=1}^{20}o_{i}-e_{i}$,
$o_{i}$ being observed fraction of $i^{th}$ amino acid and $e_{i}$
being the fraction of times it is expected to occur). Thus, for both
MSSAs (whole family and EC-specific) we have the following quantities
: functional restraint ($famF,ecF$), gap content ($famG,ecG$) and
sequence entropy ($famE,ecE$). Moreover for each MSSA, number of
sequences $<80\%$ identical to each other was taken as an indicator
of evolutionary information available in it.

\section{Benchmarking}

In order to assess the differences in residues important for whole
family and EC partition, baseline predictions were made by choosing
top-ranking residues according to whole family functional constraint
from residues which are not highly gapped ($famG<0.5$). Number of
baseline and SDR predictions is same whenever they are compared or
an overlap between them is computed. This helps in assessing whether
information in the EC-specific MSSA is distinct.

The likelihood of a residue to be an SDR is presumably proportional
to its proximity to the specific ligand. Hence, to quantify the merit
of a prediction, we defined mean proximity as the ratio of mean separation
between predicted residues and ligand. Mean relative proximity is
defined as the ratio of mean proximity to the mean separation between
all residues in the domain and the ligand. Distance between a residue
and ligand is taken to be the closest distance between residue sidechain
(mainchain for glycine) and ligand atoms. Smaller the mean relative
proximity, better the prediction. Prediction quality will also depend
on the number of distinct homologous sequences available. In case
of multiple ligands close to a domain, a residue's proximity to the
ligand is calculated with respect to the closest ligand. The basis
for SDR prediction is that it be sufficiently distinct between whole
family and EC-specific MSSAs. As \citet{funshiftakker} describe it,
an SDR should be a rate-shifted or conservation-shifted site. Additionally,
SDR should be sufficiently functionally constrained from ESSTs perspective
($ecF$). For a residue with low entropy in EC MSSA, if change in
entropy $dE$ (family MSSA sequence entropy - EC MSSA sequence entropy)
is high, it indicates that it could be SDR. Since each MSSA will be
different in its variability, it is not advisable to use same functional
constraint cutoff or entropy cutoff for all of them. This immediately
suggests two 2-step approaches : choose top $N1$ residues with high
dierence in sequence entropy between whole and EC MSSAs, then select
top $N2$ according to functional constraint in EC MSSA and vice versa.
But there could be a third and more attractive approach that combines
functional constraint from EC MSSA and sequence entropy difference.
We pursue the third approach.

We assume that SDR score of a residue is a linear combination of its
functional constraint, entropy and change in entropy, given that the
residue passes certain quality checks ($ecF>0.5$, $ecG<0.5$, $ecE<1$,
$dE>0.5$):

$SDRscore=ecF+a*(famE-ecE)-b*ecE$

In order to optimize the parameters $a,b$ and test the optimal ones,
we created a high quality test set from our examples, consisting of
23 examples drawn from SCOP families with at least 2 EC groups, each
with at $>10$ distict sequence homologs from ENZYME database. Parameters
$a,b$ were varied from 0 to 5 in steps of $0.2$ and 10 SDR predictions
were made. For each value of $a$ and $b$, SDR and baseline predictions
are made, each consisiting of 10 residues. Note that baseline predictions
are not affected by values of $a,b$. Optimization can be done with
two objectives, either to minimize the mean proximity or to maximize
the number of close ($<$\Ang{6}) residues. $a,b$ values of $0.4,1.2$
minimize the prior obective to \Ang{9.24} and yield $3.6$ close
residues per prediction, whereas $0,0.8$ maximize the latter to $4.08$
residues while yielding \Ang{9.36} for the prior. Performance of
these two $a,b$ values on different sets of examples is shown in
Table \ref{evolABperf}.

\begin{table}

\caption{Optimal values of a and b for various levels of evolutionary information
available.}

\begin{center}\begin{tabular}{|c|p{1in}|p{1in}|p{1in}|p{1in}|}
\hline 
Criteria for&
\multicolumn{2}{c|}{Mean proximity}&
\multicolumn{2}{c|}{\#close (<\Ang{6}) residues}\tabularnewline
choice of examples&
(0,0.8)&
(0.4,1.2)&
(0.0.8)&
(0.4,1.2)\tabularnewline
\hline
\hline 
>5 homologs&
10.84&
11.24&
3.35&
3.01\tabularnewline
(67 examples)&
&
&
&
\tabularnewline
\hline 
>10 homologs&
10.41&
10.64&
3.45&
3.2\tabularnewline
(55 examples)&
&
&
&
\tabularnewline
\hline 
>10 homologs, >1 EC&
9.36&
9.24&
4.08&
3.6\tabularnewline
(23 examples)&
&
&
&
\tabularnewline
\hline
\end{tabular}\end{center}

\label{evolABperf}
\end{table}

This suggests that optimal $a,b$ parameters are $0,0.8$. It is surprising
that there is no importance for the value of $dE=famE-genE$ in SDR
score. Perhaps this is due to the quality checks applied prior to
calculation of SDR scores, which demand $dE>0.5$.

\begin{figure}

\caption{Frequency of observing a certain mean proximity of SDR predictions
(binned in \Ang{1} bins) for different qualities of evolutionary
information available.}

\begin{center}\includegraphics[%
  width=150mm]{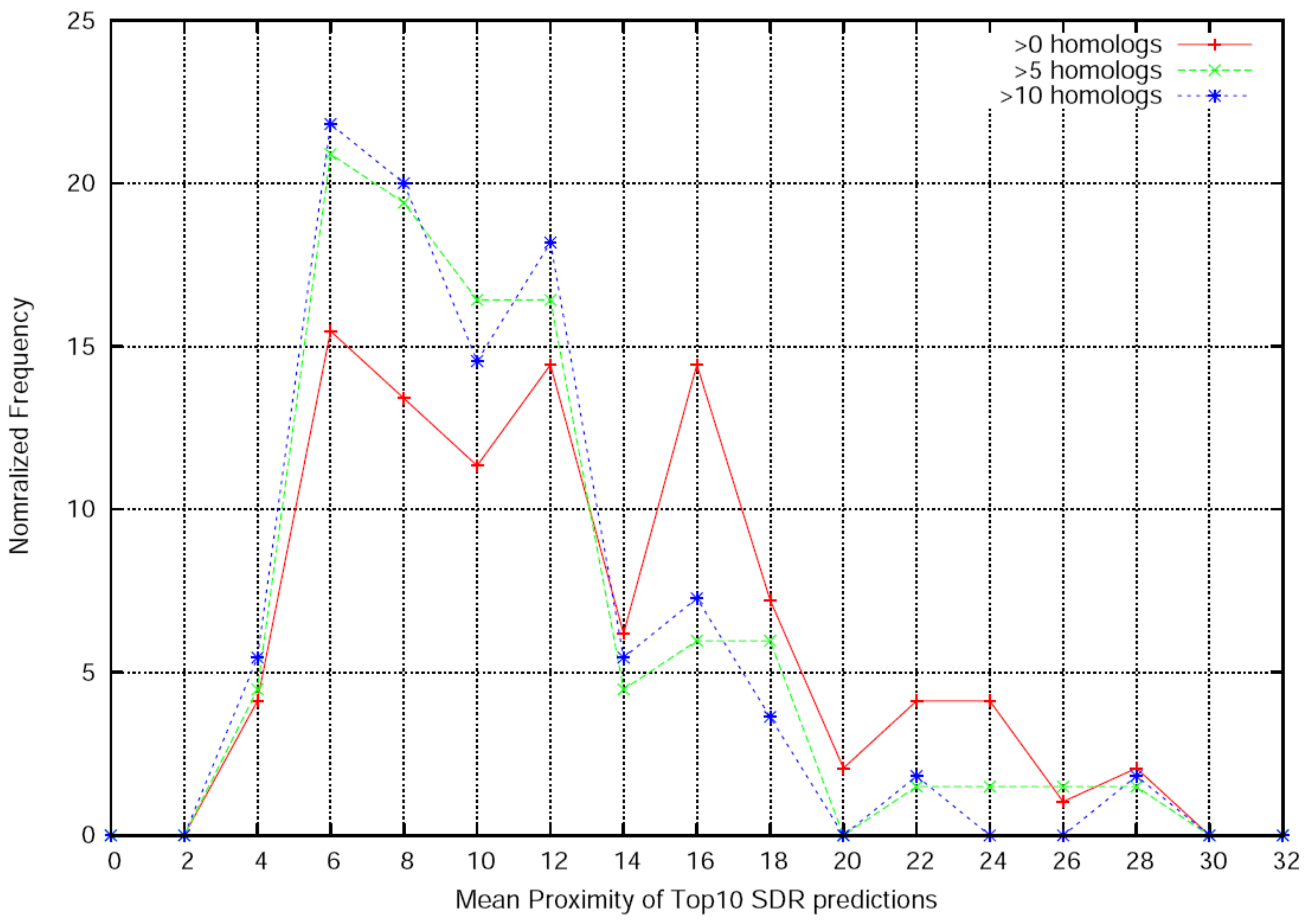}\end{center}

\label{proxDistrib}
\end{figure}

Fig.\ref{proxDistrib} shows the distribution of mean proximity in
various sets derived according to number of distinct homologs in ENZYME.
This shows that quality of evolutionary information available has
great impact on quality of predictions.

Mean relative proximity indicates how far from random is the prediction.
Table \ref{meanRelProxTable} shows that mean relative proximity depends
on quality of evolutionary information and is far from random for
both SDR and baseline predictions.

\begin{table}

\caption{Mean relative proximity in various datasets made according to number
of available distinct homologs.}

\begin{center}\begin{tabular}{|c|c|c|c|}
\hline 
Dataset&
Mean Rel. Prox.&
Mean Rel. Prox.&
Frequency of\tabularnewline
&
&
&
MRP(SDR) $\leq$ MRP(baseline)\tabularnewline
\hline 
>0 homologs&
0.67&
0.66&
34\% (33/97)\tabularnewline
\hline 
>5 homologs&
0.57&
0.66&
60\% (40/67)\tabularnewline
\hline 
>10 homologs&
0.57&
0.62&
85\% (47/55)\tabularnewline
\hline
\end{tabular}\end{center}

\label{meanRelProxTable}
\end{table}

The fraction of SDRs present in baseline predictions is $15\%$ in
all $>0,>5,>10$ homologs classes, which suggests that SDR predictions
are fairly different than baseline. This also suggests that baseline
and SDR predictions are complementary to each other.

\section{Some examples}

When quality sequence information is available, SDR predictions are
closer to specific ligand than baseline predictions which in turn
are closer than random. Here we compare our Top10 predictions with
information from literature for some examples.

\subsection{Aminotransferases}

\begin{figure}

\caption{SDR (green) and functional residue (red) predictions for 1gex, a
HspAT. Residues predicted both as functional and specificity-conferring
are colored blue. Top left panel shows Top5 predictions, top right
panel shows Top10 predictions and bottom panel zooms in on the region
around ligand in the Top10 case.}

\begin{center}\includegraphics[%
  width=150mm]{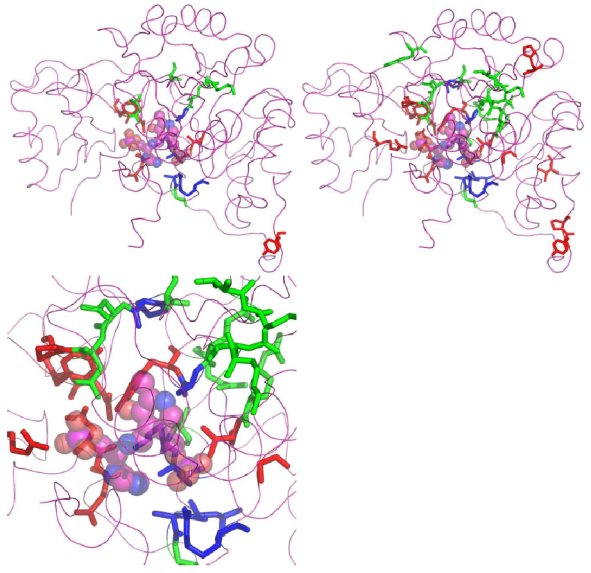}\end{center}

\label{figTransaminase}
\end{figure}

Aminotransferases or transaminases are important to amino acid biosynthesis
and unique due to their specificity to two substrates : a glutamate
and a amino-carrier. Our dataset contains two SCOP families (c.67.1.1
and c.67.1.4) that contain transaminases. Of those, we focus on SCOP
family c.67.1.1 which contains the functional categories aspartate
transaminase (AspAT, EC 2.6.1.1) and histidinol phosphate transaminase
(HspAT, EC 2.6.1.9). Other non-transaminase members of this family
include threonine adolases (EC 4.1.2.5) and alliin lyase (EC 4.4.1.4).
When Top10 predictions were analyzed in 1gex, an HspAT, we found that
SDR predictions are very well clustered around the ligands PLP and
HSP, but 5 of the 10 predictions were shared with Top10 baseline predictions.
This overlap can be attributed to degrees of functional diversity
in the SCOP family, i.e. large entropy reduction in HspAT residues
could be due to their importance to general transaminase mechanism
(as opposed to aldolase mechanism) or for substrate specificity to
histidinol phosphate (as opposed to aspartate in AspATs). In order
to increase the number of distinct predictions, Top20 baseline and
SDR predictions were used. Fig.\ref{figTransaminase} shows the predictions
for 1gexA, an HspAT from E. coli - 7 predictions are common. Catalytically
important residues (\citet{Haruyama2001}) Asn-157, Tyr-187, Lys-214
are identied as baseline, SDR and common respectively. Tyr-55, which
interacts with substrate of the other subunit, is predicted as SDR%
\footnote{This is conrmed from a similar prediction in 1gc4, an AspAT.%
}. Tyr-20, believed to be important for specificity, is not predicted
as such because it is conserved only 80\% of times, whereas a similarly
placed Tyr-55 from other subunit is much better conserved (98\% times)
and could be equally important for specificity. Ala-186, considered
important for restricting rotation of PLP's pyrimidine ring and thereby
contributing to strain essential for enzyme function, is predicted
as both SDR and baseline. Most other predicted SDRs lie close to the
substrate. Their location and AspAT counterparts suggest their role
in conferring specificty towards histidinol phosphate (see Table \ref{transaminaseTable}).

\begin{table}

\caption{Residues from speculated roles \citet{Haruyama2001} for HspAT 1gex
and how well they were predicted. The aligned residues in other subfamilies
with transaminases are also shown.}

\begin{center}\includegraphics[%
  width=150mm]{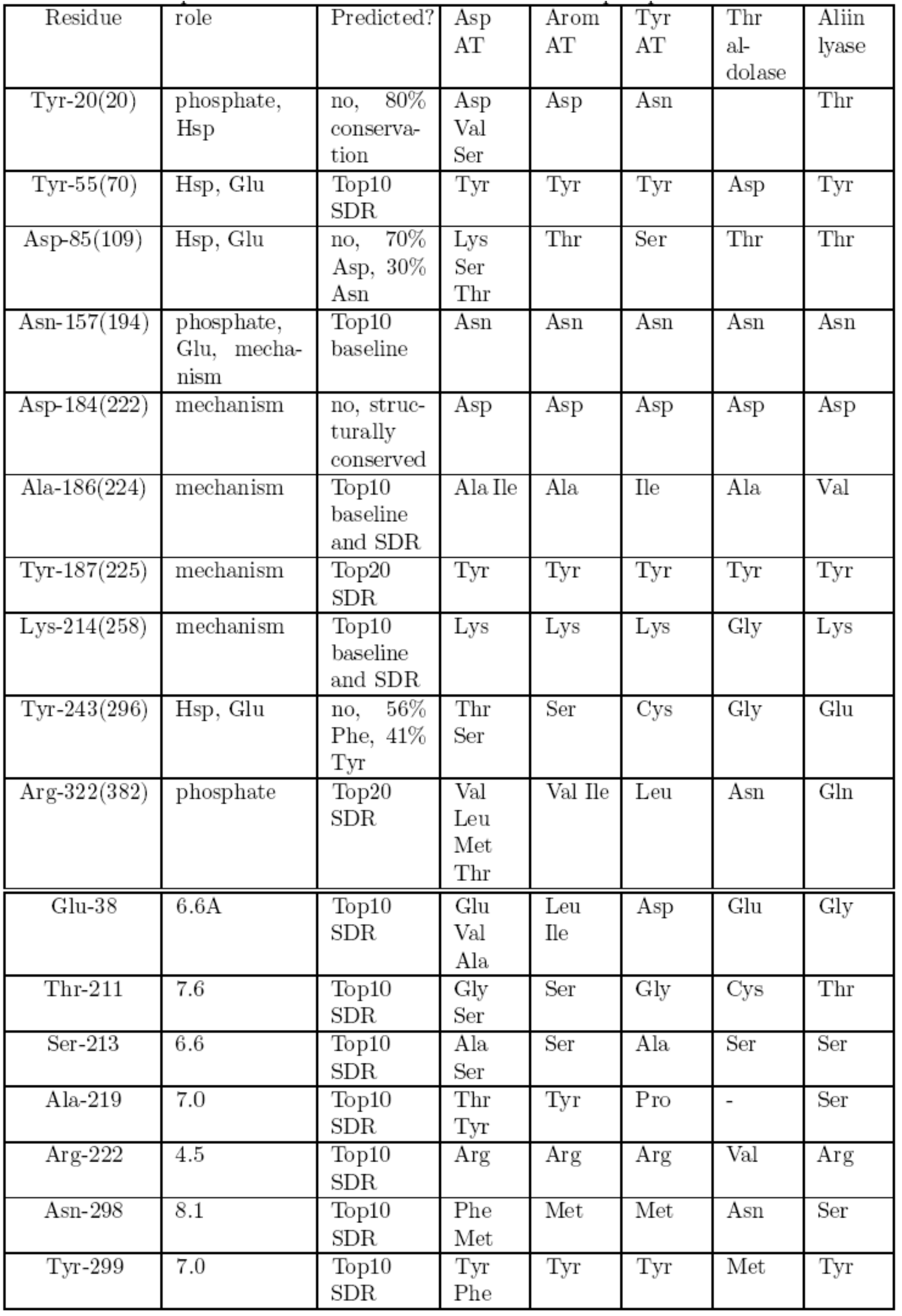}\end{center}

\label{transaminaseTable}
\end{table}

\subsection{Phosphoric monoester hydrolases}

\begin{figure}

\caption{SDR and functional residue predictions for 1cnq, a FBPase. Residue-coloring
scheme same as Fig.\ref{figTransaminase}. The bottom panel is a closer
view of the region around ligand in the top panel.}

\begin{center}\includegraphics[%
  width=100mm]{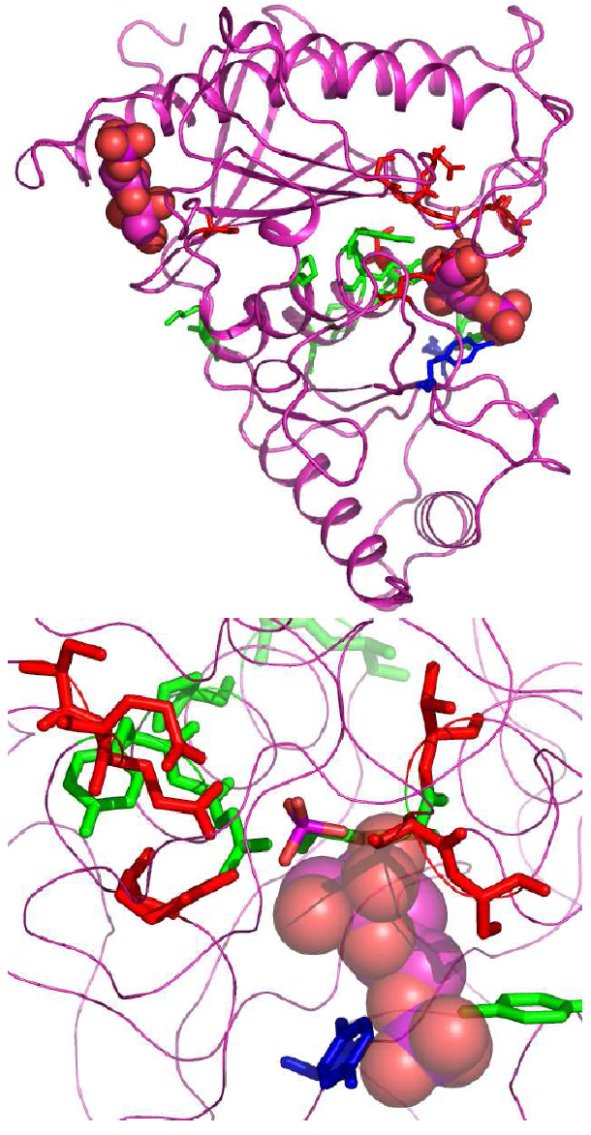}\end{center}

\label{figFBPase}
\end{figure}

SCOP family e.7.1.1 in our dataset contains 4 classes of phosphoric
monoester hydrolases, 3'(2'),5'-bisphosphate nucleotidase (EC 3.1.3.7),
Fructose-bisphosphatase (EC 3.1.3.11), Inositolphosphate phosphatase
(EC 3.1.3.25) and Inositol-1,4-bisphosphate 1-phosphatase (EC 3.1.3.57).
Here we look at the SDR and baseline predictions for 1cnq, a member
of FBPase category. FBPases are of key importance to regulation of
gluconeogenic pathway and catalyze the hydrolysis of fructose 1,6-biphosphate
to fructose 6-phosphate. They are metal dependent and are allosterically
controlled by AMP which triggers a conformational change and masks
the fructose active site. Fig.\ref{figFBPase} shows the Top10 baseline
and general predictions, the overlap in this case of 2 residues. F6P
molecule around which most predictions are clustered lies in the active
site whereas the other F6P molecule is similarly located as AMP (from
comparison with PDB 1yyz). Baseline predictions Tyr-279, Glu-280,
Tyr-244, Met-244 and common prediction Tyr-264 are within interacting
distance of F6P ligand in the active site. Most predicted SDRs form
the active site walls and differ between FBPase and IMPase (1awb)
: Arg-276 to His, Ser-96 to Gly, Ser-123 to Thr, Ser-124 to Thr (see
Table \ref{FBPaseTable}). It is surprising to see that the allosteric
site is only mildly detected. Predictions Ala-161 (Top10 SDR), Lys-290
(Top10 baseline) and Val-178 (Top20 SDR) are close and suggestive
of some role in AMP binding.

\begin{table}

\caption{Speculated roles of residues in FBPase for 1cnq from literature and
how well they were predicted. Aligned residues in other subfamilies
of hydrolases are also shown.}

\begin{center}\includegraphics[%
  width=150mm]{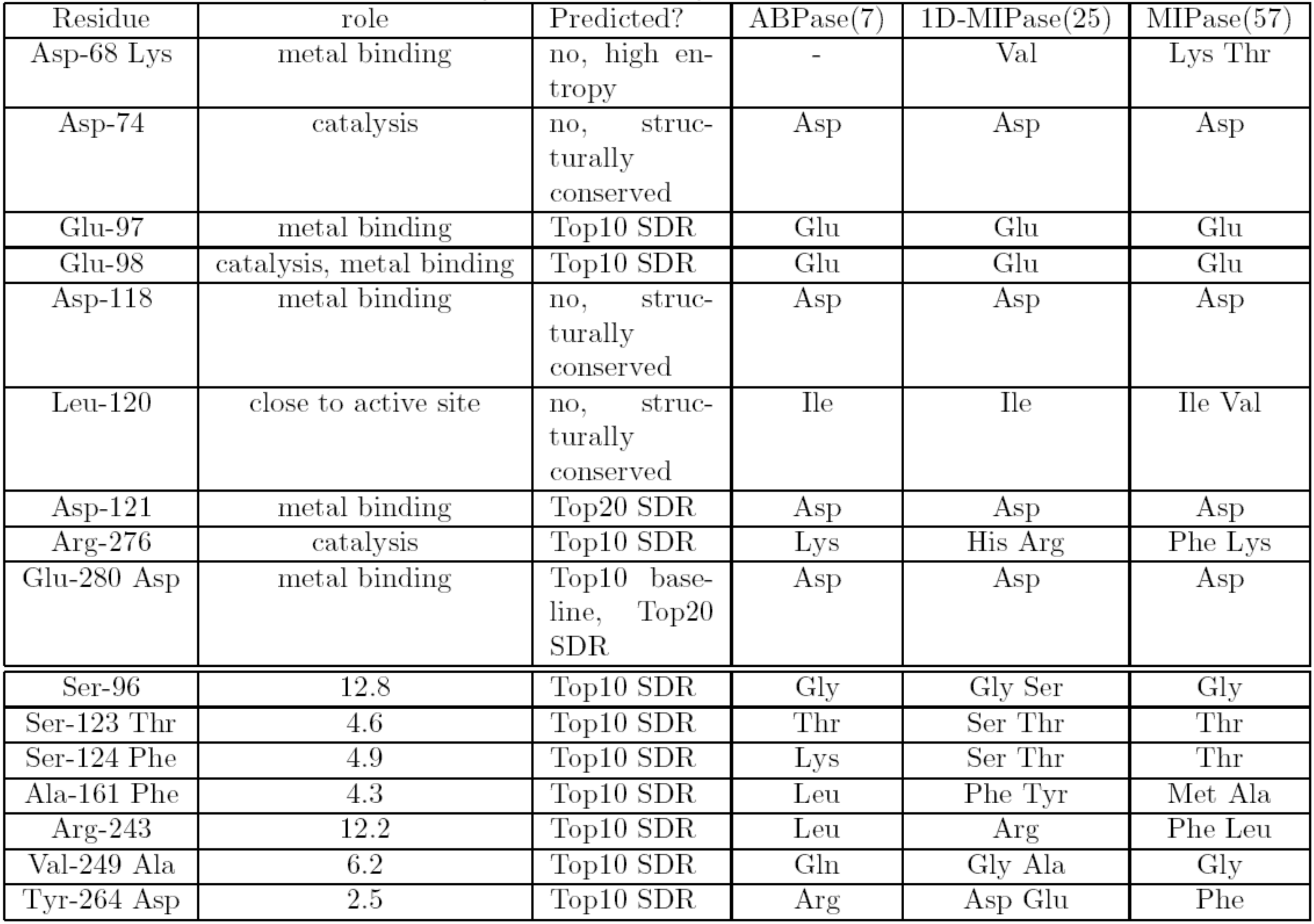}\end{center}

\label{FBPaseTable}
\end{table}

\subsection{Dehydrogenases}

\begin{figure}

\caption{SDR and functional residue predictions for 1f0y, a HAD. Residue-coloring
scheme same as Fig.\ref{figTransaminase}.}

\begin{center}\includegraphics[%
  width=100mm]{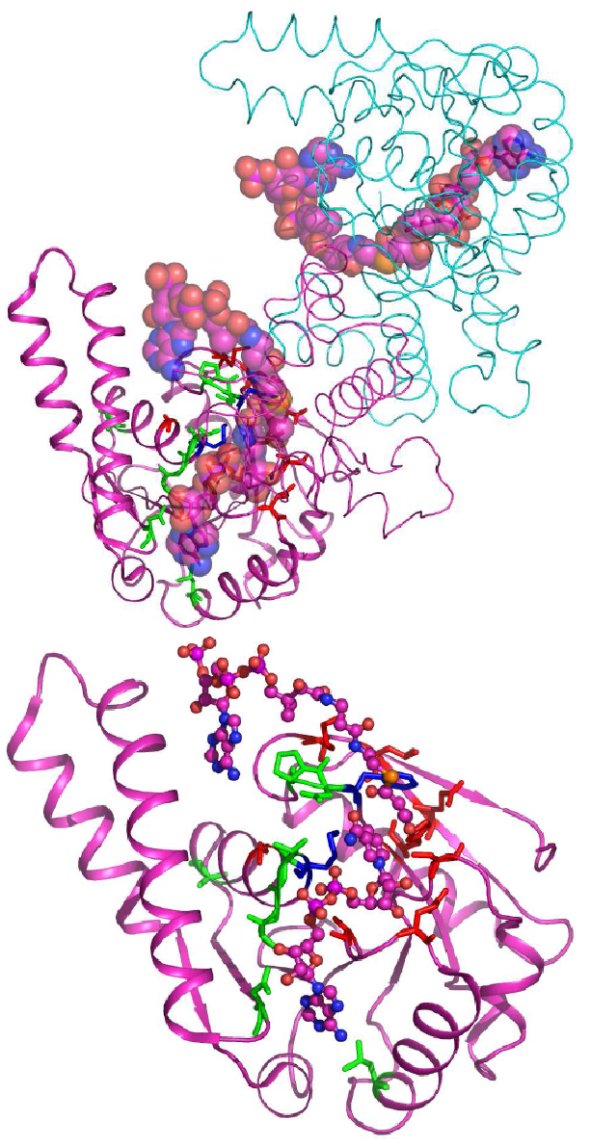}\end{center}

\label{figHAD}
\end{figure}

L-3-hydroxyacyl-CoA dehydrogenase (HAD, EC 1.1.1.35) is penultimate
enzyme in -oxidation spiral and catalyzes conversion of hydroxy group
to keto group while converting NAD+ to NADH. It consists of NAD-binding
and C-terminal domains, which undergo relative movement between NAD
binding and substrate binding events (\citet{activesiteSequestration}).
Its SCOP family is c.2.1.6, other members of which are other NAD/NADP-dependent
dehydrogenases (ECs 1.1.1.8, 1.1.1.22, 1.1.1.44). HAD is represented
in our dataset by NAD-binding domain of 1f0y (residues from A-12 to
A-203). Fig.\ref{figHAD} shows Top10 baseline and SDR predictions.
Catalytically important pair of Glu-170 and His-158 is identied as
SDRs. Ser-137, interesting due to its contact with substrate as well
as NAD, is also identied as SDR. With the exceptions of Leu-122, Ala-35
(baseline) and Gly-29, Ala-107 (SDR), all other predictions are within
interacting distance of either NAD or substrate. Ser-61 and Lys-68
are not detected due to their high entropy.

\subsection{Tryptophan biosynthesis enzymes}

\begin{figure}

\caption{SDR and functional residue predictions for TrpF. Residue-coloring
scheme same as Fig.\ref{figTransaminase}.}

\begin{center}\includegraphics[%
  width=100mm]{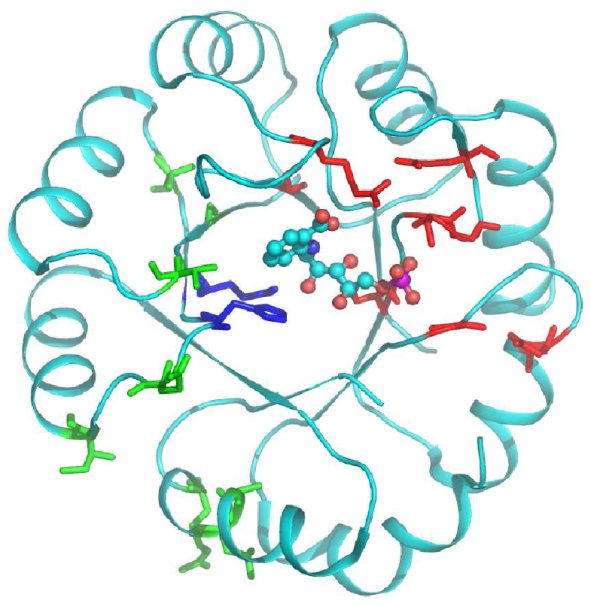}\end{center}

\label{figTRPF}
\end{figure}

Phosphoribosylanthranilate (PRA) isomerase (TrpF) is a $(\beta\alpha)_{8}$
barrel enzyme which is the most common fold adopted by enzymes and
popular among non-enzymes. TrpF (EC 5.3.1.24) shares its SCOP family
(c.1.2.4) with indole-3-glycerol-phosphate synthase (EC 4.1.1.48)
and tryptophan synthase (EC 4.2.1.20), which are all involved in Trp
biosynthesis. Top10 baseline and SDR predictions are show in Fig.\ref{figTRPF}.
His-83 and Arg-36, considered important for catalysis, are predicted.
Gln-81 (Glu in Trp synthase 1kfc), predicted as baseline and SDR,
could be important for catalysis due to its location. A few baseline
predictions are far from active site and their conservation suggests
protein-protein binding interface. Predicted SDRs lie close to ligand
and are either replaced by other residues in Trp synthase (Arg-36
to Asn) or deleted (Gln-184, Asp-178), which suggests that they could
be specificity determining.

\subsection{tRNA synthetases}

\begin{figure}

\caption{SDR and functional residue predictions for 1h3e (tyrosil tRNA synthetase).
Residue-coloring scheme same as Fig.\ref{figTransaminase}.}

\begin{center}\includegraphics[%
  width=100mm]{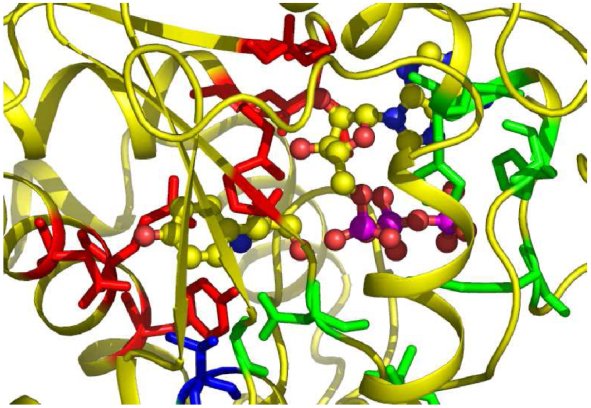}\end{center}

\label{figTyrTRNA}
\end{figure}

Aminoacyl-tRNA synthetases catalyze the process of attaching an amino
acid to its tRNA carrier so that it can be incorporated into a protein.
SCOP family c.26.1.1 contains tyrosyl-tRNA synthetase (EC 6.1.1.1)
along with other (Trp-, Glu-, Gln-) tRNA synthetases. Fig.\ref{figTyrTRNA}
shows baseline and SDR predictions for tyrosyl-tRNA synthetase 1h3e
from a thermophilic baterium T. thermophilus (\citet{tyrTRNAclass12}).
Residues important for catalysis from 51-HIGH and 233-KMSKS regions
are predicted as baseline (His-52, Gly-54, His-55, Lys-235). Predicted
SDRs lie close to the substrate and cofactor. Residues specific for
L-tyrosine binding, according to \citet{tyrTRNAspecificity} (e.g.
Thr-80, Tyr-175, Gln-179, Asp-182, Glu-197), are detected. Note that
substrate similarity makes 2 broad divisions in this family corresponding
to Trp/Tyr and Glu/Gln, each of which is subdivided into finer groups.
Table \ref{tRnaTable} shows residues structurally aligned to SDRs
in these tRNA synthetases.

\begin{table}

\caption{Residues in other tRNA synthetases aligned to predicted SDRs in tyrosil
tRNA synthetase.}

\begin{center}\includegraphics[%
  width=150mm]{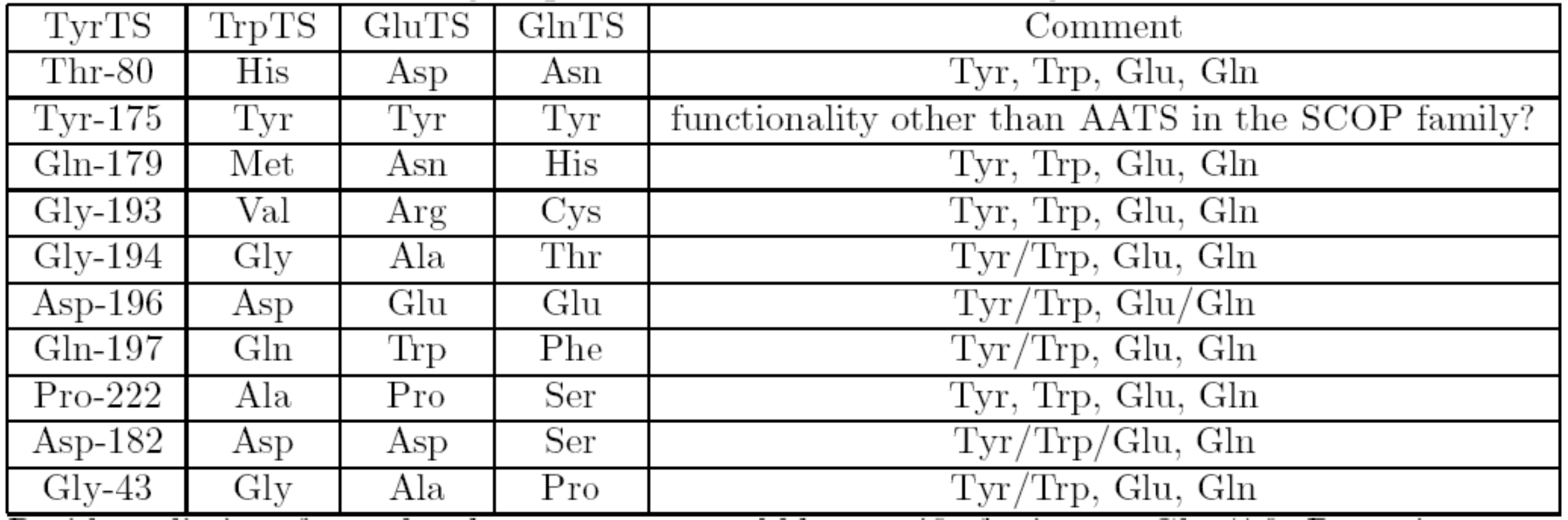}\end{center}

\label{tRnaTable}
\end{table}

Residues distinct for each substrate-group could be specific for it,
e.g. Gln-179. Detection of residue Tyr-175 as SDR suggests that there
could be more functions associated with this structural family than
these four AATSs. Detection of residues close to cofactor indicates
different/no cofactors used by other functions of this structural
family. Some residues speculated by \citet{tyrTRNAspecificity} to
be functional, stay undetected, e.g. Asn-128 which is not predicted
due to high entropy (Ser dominates the MSSA column, not Asn).

\section{Conclusion}

We have combined structural and sequence information, functional annnotation,
residue entropy and environment specific substitution tables to predict
specificity determining residues. We tested the predictions by using
information of specific ligands and in some cases, published literature.
We found that the predictions are far from random and functionally
relevant, which suggests that our approach is effective. Predictions
obtained with functional annotation (SDRs) and without it (baseline)
are different, suggesting that available functional annotation is
valuable. SDR and baseline predictions are complementary because they
enlarge the set of functionally significant residues that can be computationally
identified. We expected and found that our method cannot identify
significant residues in absence of high quality evolutionary information,
hence the importance of identifying chemically interesting patches
remains undiminished. A major concern is how to obtain functional
partitions in absence of annotation, which is similar as establishing
ortho/paralogy relationships. We plan to explore structure-sequence
scoring schemes that would help establish functional partitions reliably.
Alternatively, it would be useful to analyze the effects of constructing
a functional partition based on sequence identity. We plan to use
residue proximity information and residue contact conservation to
detect clusters which may not be conserved in the obvious sense. We
expect that cluster identification will alleviate the problem of not
identifying structurally conserved residues. The most important purpose
of SDR and catalytic residue identification is to help classify SNPs
into normal/deleterious classes and this would be an important avenue
to explore in near future.

\subsection*{Acknowledgements}

We thank Dr Kenji Mizuguchi and Dr Vijayalakshmi Chelliah for helpful
discussions. Swanand Gore thanks Cambridge Commonwealth Trust and
Universities UK Overseas Research Studentship for funding.

\bibliographystyle{marko}
\bibliography{sdr}

\end{document}